\documentclass{article}

    \usepackage[final]{neurips_2020}

\usepackage[utf8]{inputenc} 
\usepackage[T1]{fontenc}    
\usepackage{hyperref}       
\usepackage{url}            
\usepackage{booktabs}       
\usepackage{amsfonts}       
\usepackage{nicefrac}       
\usepackage{microtype}      
\usepackage{graphicx}
\usepackage{float}
\usepackage{multirow}
\usepackage{amssymb}
\usepackage{amsmath,amscd}
\usepackage{amsthm}
\usepackage{cancel}     
\usepackage{centernot}  
\usepackage{graphicx}   
\usepackage{soul}       
\usepackage{color}      
\usepackage{enumitem}   
\usepackage{commath}    
\usepackage{subcaption} 

\usepackage{xcolor}

\newcommand{\jpcomment}[1]{\textcolor{blue}{#1 (JP)}}
\newcommand{\hkcomment}[1]{\textcolor{teal}{#1 (HK)}}
\newcommand{\klcomment}[1]{\textcolor{red}{#1 (KL)}}

\definecolor{mygreen}{HTML}{008000}
\newcommand{\jpedit}[1]{\textcolor{orange}{#1}}
\newcommand{\hkedit}[1]{\textcolor{mygreen}{#1}}
\newcommand{\kledit}[1]{\textcolor{magenta}{#1}}

\renewcommand{\klcomment}[1]{}
\renewcommand{\jpcomment}[1]{}
\renewcommand{\hkcomment}[1]{}
\renewcommand{\kledit}[1]{{#1}}
\renewcommand{\jpedit}[1]{{#1}}
\renewcommand{\hkedit}[1]{{#1}}

\title{A Correspondence Variational Autoencoder \\ \kledit{for} Unsupervised Acoustic Word Embeddings}

\author{%
Puyuan Peng \\
  Department of Statistics\\
  University of Chicago, USA\\
  \texttt{pengpuyuan@uchicago.edu} \\
  \And
  Herman Kamper \\
  Department of Electrical and Electronic Engineering\\
  Stellenbosch University, South Africa \\
  \texttt{kamperh@sun.ac.za} \\
  \AND
  Karen Livescu \\
  Toyota Technological Institute at Chicago, USA \\
  \texttt{klivescu@ttic.edu} \\
}

\begin{document}

\maketitle

\begin{abstract}
We propose a new unsupervised model for mapping a variable-duration speech segment to a fixed-dimensional representation. The resulting \textit{acoustic word embeddings} can form the basis of search, discovery, and indexing systems for low- and zero-resource languages.  Our model, which we refer to as \hkedit{a} maximal-sampling correspondence variational autoencoder (MCVAE), is a recurrent neural network~(RNN) trained with a novel self-supervised correspondence loss that encourages consistency between embeddings of different instances of the same word.  Our training scheme improves on previous correspondence training approaches through the use \kledit{and comparison} of multiple samples from the approximate posterior distribution. 
In the zero-resource setting, the MCVAE can be trained in an unsupervised way, without any ground-truth word pairs, by using the word-like segments discovered via an unsupervised term discovery system. In both this setting and a semi-supervised low-resource setting (with a limited set of ground-truth word pairs), the MCVAE outperforms previous state-of-the-art models, such as Siamese-, CAE- and VAE-based RNNs.
\end{abstract}

\section{Introduction}
Acoustic word embeddings (AWEs) are representations of arbitrary-length speech segments in a fixed-dimensional space, allowing for easy comparison between acoustic segments \citep{levin2013fixed}. 
\kledit{AWEs have been used to improve performance in multiple}
applications, such as query-by-example speech search~\citep{settle2017query,ao2018query,Jung2019AdditionalSD,Yuan2018LearningAW}, automatic speech recognition~\citep{Bengio2014WordEF,settle2019acoustically,shi2021whole}, and zero-resource speech processing~\citep{Kamper2016UnsupervisedWS,Kamper2017segmental}.  

While supervised AWE models have shown impressive performance,
\kledit{here we} focus on unsupervised and semi-supervised settings where transcribed data is unavailable or very limited. 
This is the case for many zero- or low-resource languages, which make up most of the languages spoken in the world today~\citep{Besacier2014AutomaticSR}.
\hkedit{One of the first unsupervised AWE models was proposed by \cite{chung2016unsupervised}, who trained an encoder-decoder RNN as an autoencoder (AE) to reconstruct unlabelled speech segments.
\cite{kamper2019truly} extended this to the correspondence autoencoder RNN (CAE-RNN): \kledit{Instead} of trying to reconstruct an input segment directly, it tries to reconstruct another instance of the same class as the input.
Since labelled data isn't available in the zero-resource setting, an unsupervised term discovery (UTD) system~\citep{park2007unsupervised} is used to automatically discover word-like training pairs.}

\hkedit{In this work we extend these unsupervised models, proposing the maximal sampling correspondence variational autoencoder (MCVAE).
In contrast to the above models, this is a generative probabilistic model that can be seen as an extension of a variational autoencoder (VAE)~\citep{kingma2013auto}.
It improves on the CAE-RNN through the use and comparison of multiple samples from the approximate posterior distribution, and is trained with a novel self-supervised correspondence loss which encourages different instances of the same (discovered) word type to have similar latent embeddings.
We compare the MCVAE to previous approaches in both the unsupervised setting (where pairs from a UTD system are used for training) and the semi-supervised setting (where limited amounts of labelled data are used to provide weak supervision).
In both settings we show that it outperforms previous approaches in a word discrimination task on two languages.
}

\section{Problem formulation and existing approaches}
\jpedit{Given unlabeled acoustic segments $(x^{(1)}, x^{(2)}, ..., x^{(N)})$ and/or segment pairs $\{(x_1^{(1)}, x_2^{(1)}), (x_1^{(2)}, x_2^{(2)}), \cdots, (x_1^{(N)}, x_2^{(N)})\}$, }\hkedit{the goal of our AWE models\footnote{Other goals are possible, such as encoding semantic similarity~\citep{chung2018speech2vec}.} is to learn a}
\kledit{function} $f$ \hkedit{such that}
(a) $f(x)$ \hkedit{maps the acoustic sequence} $x$ into \hkedit{a} \kledit{fixed}-dimensional embedding\hkedit{, and} (b) $f(x_1)$ and $f(x_2)$ are close according to some distance measure (e.g.\ cosine \hkedit{distance}) 
\hkedit{if and only if} $x_1$ and $x_2$ correspond to \hkedit{instances of} the same \hkedit{linguistic type (words in our case)}.  \hkedit{Below we introduce three existing AWE models.}

\hkedit{\bf The autoencoder recurrent neural network~(AE-RNN)}
consists of \hkedit{an} encoder and \hkedit{a} decoder \hkedit{RNN~\citep{chung2016unsupervised}}.
\hkedit{The encoder} takes a variable-length acoustic segment $x$ \kledit{and} embeds it as a 
\kledit{fixed-dimensional} vector $z$, and \hkedit{the} decoder \hkedit{then uses} $z$ \hkedit{to produce} the reconstruction $\hat{x}$. The loss function for \hkedit{the} AE-RNN is \hkedit{the} empirical reconstruction error: $\frac{1}{N}\sum_{i=1}^{N}\norm{x^{(i)}-\hat{x}^{(i)}}^2$.
\hkedit{A probabilistic variant of the AE-RNN was proposed in~\cite{kamper2019truly}, where the model is trained as a variational autoencoder (VAE). We introduce and discuss this VAE-based model in more detail in Section~\ref{sec:annealVAE}, and extend it in the subsequent sections.}

\hkedit{{\bf The correspondence autoencoder RNN (CAE-RNN)}, proposed in~\citep{kamper2019truly} and further analyzed in~\citep{Matusevych2020AnalyzingAA,Kamper2020MultilingualAW}

, is} trained with a correspondence loss, $\frac{1}{N}\sum_{i=1}^{N}\norm{x_2^{(i)}-\hat{x}_1^{(i)}}^2$, where $x_1^{(i)}$ and $x_2^{(i)}$ are \hkedit{instances} of the same word type (or belong to the same \hkedit{discovered} cluster\hkedit{---see below}) \kledit{and $\hat{x}_1^{(i)}$ is the model output when the input is $x_1^{(i)}$}.
\hkedit{Here} $x_1^{(i)}$ is the input to the encoder, \hkedit{and} we want the model to reconstruct $x_2^{(i)}$ \hkedit{as its output}. 
This loss explicitly helps the model to utilize pair information.
\hkedit{One shortcoming is that}
it might be too hard a problem to reconstruct $x_2^{(i)}$ from $x_1^{(i)}$. In Section~\ref{sec:compare}, we \kledit{propose} a probabilistic version of CAE-RNN to \kledit{address}
this problem.

\hkedit{\bf The SiameseRNN} is a discriminative \hkedit{AWE} model~\hkedit{\citep{settle2016discriminative}}\hkedit{, based on the early work of~\citep{Bromley1993SignatureVU}}. It \hkedit{consists of an} 
RNN encoder which takes acoustic segments and outputs the final state \hkedit{(or a transformation of the final state)} as the embedding for the \hkedit{input} segment. To encourage embeddings of segments of the same word type to be closer in the embedding space \kledit{than embeddings of different-type segments}, \hkedit{the} SiameseRNN is trained by minimizing the triplet loss, $\frac{1}{N}\sum_{i=1}^{N}\text{max}(0, m + d_{\text{cos}}(x_a^{(i)}, x_d^{(i)})-d_{\text{cos}}(x_a^{(i)}, x_s^{(i)}))$, where $(x_a^{(i)}, x_s^{(i)})$ \hkedit{is a positive pair and} $x_d^{(i)}$ is a random \hkedit{negative} sample from the corpus.
\hkedit{Since their introduction~\citep{kamper2016deep,settle2016discriminative},} \kledit{work on Siamese network-based embeddings has further improved them and explored their}
applications~\citep{Yang2019LinguisticallyInformedTO,Yuan2018LearningAW,Lim2018LearningAW}.  
We will show in Sections\hkedit{~\ref{sec:unsup} and~\ref{sec:semi}} that the discriminative \hkedit{nature} of \hkedit{the} SiameseRNN makes it suitable for AWEs when we have high quality training pairs.

\hkedit{While the AE-RNN can be trained without any supervision, both the CAE-RNN and SiameseRNN require paired examples.
These can be obtained from transcriptions in cases where (limited amounts of) labelled speech data are available.
But in zero-resource settings, only unlabelled audio is available.
In this case, we can use an unsupervised term discovery (UTD) system~\citep{park2007unsupervised,jansen2010towards} to discover recurring word- or phrase-like patterns in the unlabelled data, thereby automatically constructing noisy training pairs.
Since the UTD system is unsupervised, the overall AWE approach} \kledit{is then unsupervised.}

\section{The maximal sampling correspondence variational autoencoder (MCVAE)}
In this section we introduce our new models, starting with an extension of the VAE. We then introduce the new CVAE and MCVAE models, which can be seen as combinations of the VAE and the CAE-RNN for obtaining AWEs.

\vspace{-.05in}
\subsection{Variational autoencoder for acoustic word embeddings}\label{sec:annealVAE}
\vspace{-.05in}
\hkedit{A VAE was first used for AWEs in~\cite{kamper2019truly}, but it performed \kledit{poorly}.
Apart from that work, the only other study to consider generative models for AWEs is that of~\cite{beguvs2020ciwgan}, who used \kledit{generative adversarial networks (GANs)}.
Here we extend the VAE approach of~\cite{kamper2019truly}.
}

\begin{figure}
  \centering
  \begin{minipage}[t]{0.475\textwidth}
      \centering
      \includegraphics[width=0.45\textwidth]{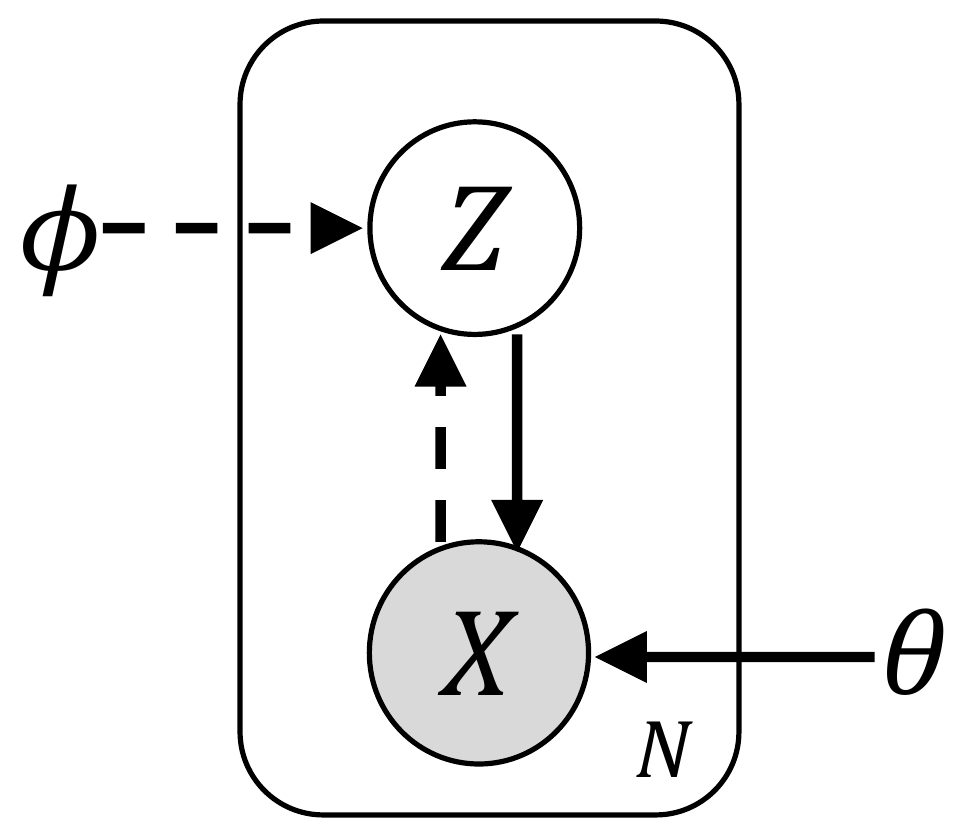} 
      \caption{Graphical \kledit{model for a} VAE. $X$ is \kledit{the} observation and $Z$ is \kledit{the} latent variable. Solid lines denote the generative model and \kledit{dashed} lines denote \kledit{the} variational approximation.}\label{fig:vaepgm}
  \end{minipage}\hfill
  \begin{minipage}[t]{0.475\textwidth}
      \centering
      \includegraphics[width=0.7\textwidth]{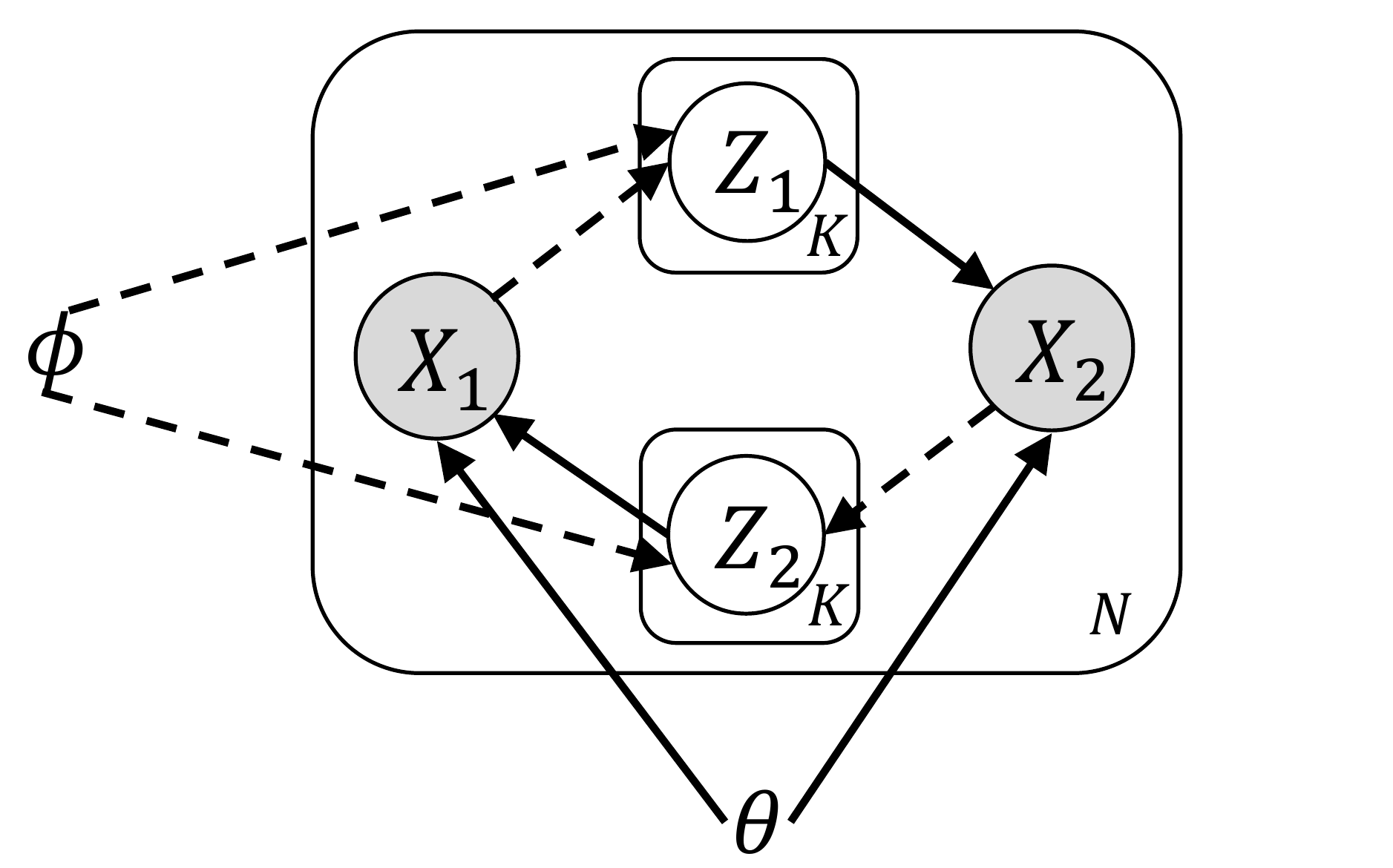} 
      \caption{Graphical \kledit{model} for the correspondence training \kledit{approach.} }\label{fig:cvaepgm}
  \end{minipage}
\end{figure}

\kledit{To establish terminology, the graphical model of a VAE}
is shown in Figure~\ref{fig:vaepgm}, where $X$ is the observed \kledit{input (here, acoustic segment)} and $Z$ is the latent \kledit{vector} variable \kledit{(here, the acoustic word embedding)}.\footnote{We use uppercase letters to denote random variables and lowercase letters to denote their realizations.  }  
Solid lines denote the generative model $p(Z)p_{\theta}(X|Z)$ and \kledit{dashed} lines denote the variational approximation $q_{\phi}(Z|X)$ to the intractable posterior $p_{\theta}(Z|X)$. 
\hkedit{The variational} 
approximation $q_{\phi}(Z|X)$ is assumed to be a diagonal Gaussian distribution whose mean and log variance is the output of \kledit{an} encoder network. 
\kledit{The likelihood} model $p_{\theta}(X|Z)$ is assumed to be \hkedit{a} spherical Gaussian, and its mean is the output of \kledit{a} decoder network.   
$\phi$ and $\theta$ are weights of the encoder and decoder networks, \kledit{respectively,} and they are jointly learned by maximizing the evidence lower bound~(ELBO):
\begin{eqnarray}\label{eq:elbo}
  \text{ELBO} = E_{Z\sim q_{\phi}(Z|x)} \log{p_{\theta}(x|Z)} - D_{KL}(q_{\phi}(Z|x)||p(Z))
\end{eqnarray}
\kledit{where $x$ is a data point (acoustic segment), \kledit{i.e.~an} instantiation of $X$.}  It can be shown that maximizing \kledit{the} ELBO is equivalent to minimizing 
$D_{KL}(q_{\phi}(Z|x)||p_{\theta}(Z|x))$~\citep{kingma2013auto}.  

\kledit{The first term in equation~\ref{eq:elbo} cannot be computed analytically, so we estimate it using samples of the latent variable;
the \jpedit{objective} over a training set $(x^{(1)}, x^{(2)}, ..., x^{(N)})$ then becomes:}
\begin{align}\label{eq:mcelbo}
  \kledit{J_{\text{VAE}}} = \frac{1}{N}\sum_{i = 1}^{N}\left\{\frac{1}{K}\sum_{k=1}^{K} \log{p_{\theta}(x^{(i)}|z^{(k)})}- D_{KL}(q_{\phi}(Z|x^{(i)})||p(Z)) \right\}
\end{align}
\jpedit{where $z^{(k)} \overset{\text{i.i.d}}{\sim} q_{\phi}(\jpedit{Z}|x^{(i)})$, with $k=1,2,\cdots,K$.}

For the building blocks of the encoder-decoder network, we \hkedit{use} 
\kledit{multi-layer RNNs}
with GRU~\citep{Cho2014LearningPR} \kledit{units}. Note that other layers such as convolutional or transformer layers~\citep{Vaswani2017AttentionIA} \hkedit{could} 
also be used.

One \kledit{well-known issue with VAEs is posterior collapse}~\citep{razavi2019preventing}, \kledit{where}
the decoder learns to ignore the latent variable $Z$ and the approximate posterior 
\kledit{collapses} to the prior. 
Preventing the posterior from collapsing has been a very active research \kledit{question~\citep{He2019LaggingIN,razavi2019preventing,Lucas2019DontBT}}.
Here we adopt the conditioning scheme used in \cite{kamper2019truly} and the KL term annealing trick used in \cite{Bowman2016GeneratingSF} to tackle the posterior collapse problem. 

\begin{figure}
\vspace{-.2in}
  \begin{center}
    \includegraphics[width=0.8\textwidth]{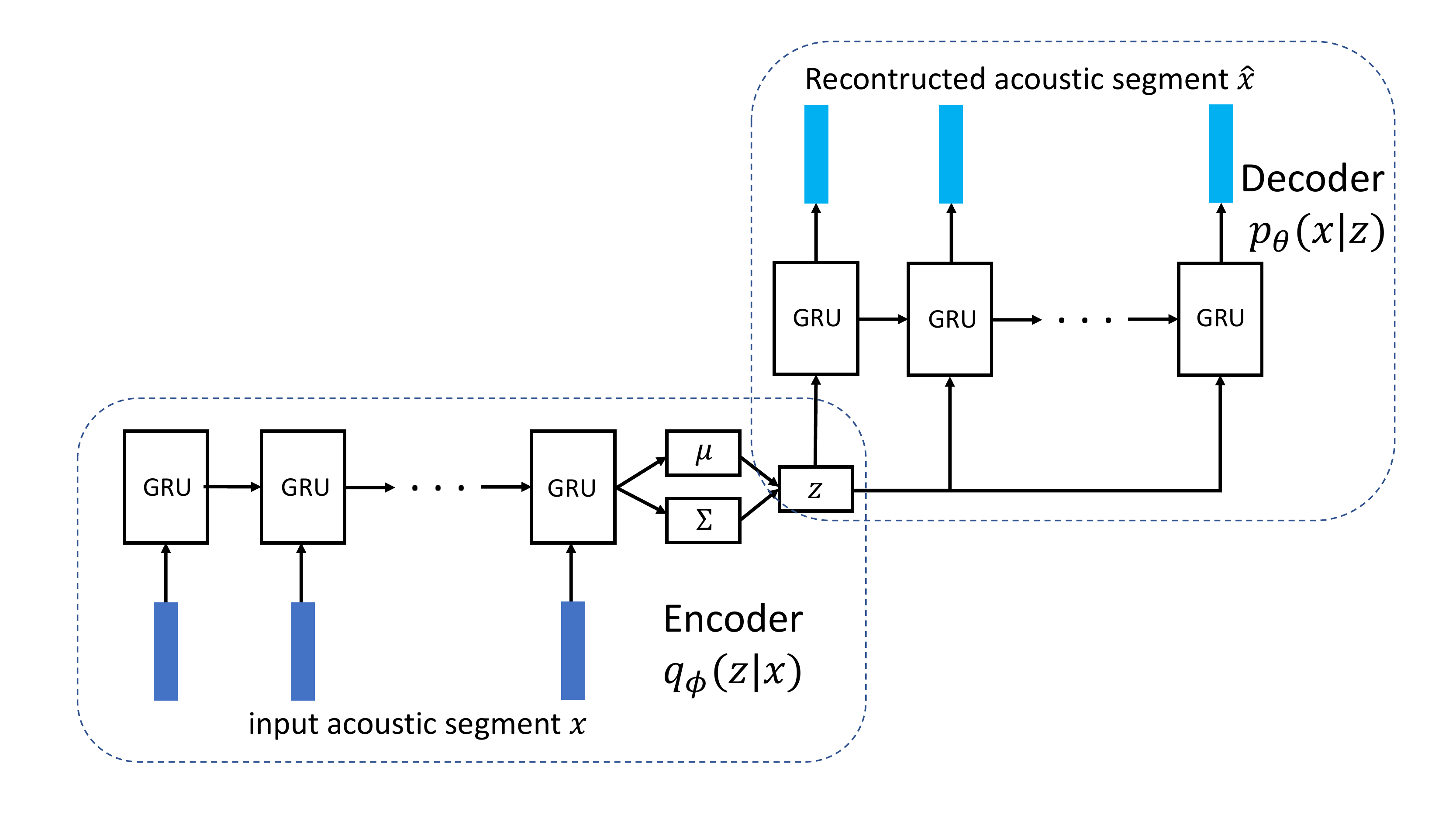}
    \vspace{-.15in}
    \caption{Network architecture of our model.}\label{fig:rnnvae}
  \end{center}
  \vspace{-.25in}
\end{figure}

\textbf{Conditioning scheme.} We use \jpedit{samples of} $Z$ as the input of the decoder RNN at every time step, \kledit{as}
shown in Figure~\ref{fig:rnnvae}. \kledit{Note that the network can be a bidirectional RNN with multiple GRU layers,} in which case $\mu$ and $\Sigma$ will be a concatenation of the final states of the last forward and backward layer. We also tried the conditioning scheme in \cite{Bowman2016GeneratingSF}, which uses the latent variable as initial state for the decoder and \kledit{the target output from the previous step in a scheduled sampling scheme, to encourage the decoder to rely more on the initial state.}
However, in our experiments, 
\hkedit{this approach}
fails to learn a competitive representation, \kledit{
as} measured by our discrimination task.

\textbf{KL term annealing.} We give a weight to the KL term in our objective function \kledit{(equation~\ref{eq:mcelbo})} and gradually increase it from $0$ to $1$ as training proceeds. \kledit{Similarly} to~\cite{Bowman2016GeneratingSF}, we \kledit{use} a sigmoid function for the weight:
\begin{equation}
  \text{weight} = \frac{1}{1+e^{-k(t-s_0)}},
\end{equation}

\jpedit{where $t$ is the training iteration, and} $s_0$ and $k$ are hyperparameters that affect the starting point of the annealing and the 
\kledit{increment} in weight at each step during annealing.  
\vspace{-.1in}
\subsection{A self-supervised correspondence objective}\label{sec:compare}
\vspace{-.05in}
\kledit{After training the VAE, it}
can infer the posterior of \kledit{the} latent variable $Z$ and \kledit{sample} from the posterior to reconstruct $X$. 
If speech segments 
\kledit{$x_1$ and $x_2$} correspond to the same word type (or belong to the same cluster in \kledit{the} unsupervised setting), the approximate posteriors 
\kledit{$p_{\phi}(Z|x_1)$ and $p_{\phi}(Z|x_2)$}
should be similar.
\kledit{We next ask the question:}
if we know \kledit{that $x_1$ and $x_2$} correspond to the same word type, how can we explicitly express this information and use it to further \kledit{improve} the model?
\vspace{-.05in}
\paragraph{Correspondence VAE.}  Inspired by \cite{kamper2019truly}, we propose a probabilistic correspondence training \kledit{approach}, \kledit{whose graphical model is shown in Figure~\ref{fig:cvaepgm}.}  
\kledit{As in} Figure~\ref{fig:vaepgm}, solid lines correspond to \kledit{the} generative model and \kledit{dashed} lines correspond to \kledit{the} inference model.  To solve for parameters $\theta$ and $\phi$ in the graphical model via optimization, a natural choice \kledit{for} the objective function is 
\begin{eqnarray}\label{eq:CoKL}
  D_{KL}(q_{\phi}(Z|x_1)||p_{\theta}(Z|x_2))+D_{KL}(q_{\phi}(Z|x_2)||p_{\theta}(Z|x_1)).
\end{eqnarray}
This objective expresses that we want the posterior conditioned on $x_1$ to be close to the posterior conditioned on $x_2$.  Minimizing this loss is equivalent to maximizing
\begin{eqnarray}
  \text{ELBO}_1 + \text{ELBO}_2 = E_{Z\sim q_{\phi}(Z|x_1)} \log{p_{\theta}(x_2|Z)} - D_{KL}(q_{\phi}(Z|x_1)||p(Z))\nonumber\\+E_{Z\sim q_{\phi}(Z|x_2)} \log{p_{\theta}(x_1|Z)} - D_{KL}(q_{\phi}(Z|x_2)||p(Z))  \nonumber
\end{eqnarray}
Given data $\{(x_1^{(1)}, x_2^{(1)}), (x_1^{(2)}, x_2^{(2)}), \cdots, (x_1^{(N)}, x_2^{(N)})\}$, the objective can be approximated
\begin{eqnarray}\label{eq:CoELBO}
  J_{\text{CVAE}} = \frac{1}{N}\sum_{i=1}^{N}\left\{\left[\frac{1}{K}\sum_{k_2=1}^{K}\log p_{\theta}(x_1^{(i)}|z_2^{(k_2)})\right] - D_{KL}(q_{\phi}(Z|x_1^{(i)})||p(Z))\right.\nonumber \\ \left. + \left[\frac{1}{K}\sum_{k_1=1}^{K}\log p_{\theta}(x_2^{(i)}|z_1^{(k_1)})\right]- D_{KL}(q_{\phi}(Z|x_2^{(i)})||p(Z)) \right\}, \nonumber\\ z_1^{(1)}, \cdots, z_1^{(K)} \overset{\text{i.i.d}}{\sim} q_{\phi}(Z|x_1^{(i)}), \quad z_2^{(1)}, \cdots, z_2^{(K)} \overset{\text{i.i.d}}{\sim} q_{\phi}(Z|x_2^{(i)}).
\end{eqnarray}
We \kledit{refer to} the model trained with \kledit{this} \hkedit{new} objective function as \kledit{a correspondence VAE, or} CVAE.\footnote{Note that this is not to be confused with \hkedit{the} \kledit{conditional} VAE.}

\jpedit{In practice, a pair of samples of the same word can be very different, making the task of the CVAE too challenging. We next propose a technique intended to better handle the gap between different acoustic segments of the same word type.}

\vspace{-.05in}
\paragraph{Maximal sampling correspondence VAE.}  Given \kledit{the} data pair $(x_1, x_2)$, \kledit{we} pass $x_1$ to the encoder network and \kledit{get the} approximate posterior $q_{\phi}(Z|x_1)$. \kledit{We} then 
\kledit{draw independent samples} $z_1, z_2, ..., z_K$ from it 
\kledit{and require} $x_2$ to 
\kledit{be likely only according to} the ``best'' $p_{\theta}(X|z_k)$ and vice versa.
\kledit{In other words,} given speech segment pairs $\{(x_1^{(1)}, x_2^{(1)}), (x_1^{(2)}, x_2^{(2)}), \cdots, (x_1^{(N)}, x_2^{(N)})\}$, the objective function is
\begin{eqnarray}\label{eq:MCVAE}
  J_{\text{MCVAE}} = \frac{1}{N}\sum_{i=1}^{N}\left\{\left[\max_{k_2}\log p_{\theta}(x_1^{(i)}|z_2^{(k_2)})\right] - D_{KL}(q_{\phi}(Z|x_1^{(i)})||p(Z))\right.\nonumber \\ \left. + \left[\max_{k_1}\log p_{\theta}(x_2^{(i)}|z_1^{(k_1)})\right]- D_{KL}(q_{\phi}(Z|x_2^{(i)})||p(Z)) \right\}, \nonumber\\ z_1^{(1)}, \cdots, z_1^{(K)} \overset{\text{i.i.d}}{\sim} q_{\phi}(Z|x_1^{(i)}), \quad z_2^{(1)}, \cdots, z_2^{(K)} \overset{\text{i.i.d}}{\sim} q_{\phi}(Z|x_2^{(i)}).
\end{eqnarray}
Since we are \kledit{maximizing} over samples, we \kledit{refer to this technique as} ``maximal sampling'' and \hkedit{to the} 
model \hkedit{as the maximal sampling correspondence VAE,} 
or MCVAE. 

Note that the number of samples $K$ has different roles in \hkedit{the equations above.}
In \hkedit{equations}~\ref{eq:mcelbo} and \ref{eq:CoELBO},  a larger $K$ can reduce the variance of the 
estimator and therefore stabilize training. 
In \kledit{equation~\ref{eq:MCVAE}}, $K$ 
\kledit{represents} how much the embedding space should be explored. 
If we assume that $x_1$ and $x_2$ are \kledit{close,}
then a small $K$ is preferable, since we don't need (and don't want) too much variation when \hkedit{reconstructing} 
$x_2$ from $x_1$. 
On the other hand, if we \hkedit{expect}
$x_1$ and $x_2$ \hkedit{to be far apart,} 
we might need a larger $K$. In our experiments, $K$ is tuned as a hyperparameter on \kledit{a} validation set. In practice we put a weight on the KL term in equation~\ref{eq:MCVAE}, a common practice in the VAE literature which can be viewed as a way to balance the quality of the reconstruction and simplicity of the representation \citep{Higgins2017betaVAELB,Chorowski2019UnsupervisedSR}. In addition, 
the weight on the KL term also controls the latent space we want our model to explore:  we find that a large weight on the KL term will lead to a large posterior variance and thus 
more variable reconstructions. Batch size is also an important factor,
since the model is optimized using stochastic gradient descent; if the batch size is too small, it's possible that within one batch there are only two instances (i.e. one pair) for many word types, and this will encourage deterministic mapping between the two instances and therefore make the optimization difficult.

\vspace{-.05in}
\section{Experiments}\label{sec:exper}
\vspace{-.05in}
\subsection{Experimental setup}\label{sec:setup}
While 
\kledit{AWEs can be used for a variety of} downstream tasks, for comparison \kledit{purposes we use a word discrimination (same-different) task~\citep{Carlin2011RapidEO}, often used in prior work} to evaluate the quality of embeddings. 
In the same-different task, given a pair of speech segments \kledit{corresponding to one word each}, we must 
\kledit{determine whether these segments are examples of the same word}. 
\kledit{For each pair of segments in the test set, we compute the cosine distance
between their embeddings.}
Two segments can then be classified as being of the same or different type based on 
\kledit{a threshold on the cosine distance}, and a precision-recall curve is obtained by varying the threshold. 
The area under this curve, \kledit{referred to as the average precision (AP),} is used as \kledit{the} final evaluation metric.
\kledit{AP ranges from $0$ to $1$ and \hkedit{we} report it as a percentage.}

\kledit{For all experiments,
the encoder} and decoder networks are both 2-layer bidirectional GRU \kledit{networks}, with $300$ hidden units.  The final states of the second forward and backward layers are concatenated and compressed to a \kledit{$260$-dimensional} vector via a linear layer.  \kledit{This vector is split in half, with the mean and variance of $q_{\phi}(Z|x)$ each forming a half, and the final embedding is the $130$-dimensional mean vector} (which is the same \kledit{dimensionality as in} \cite{kamper2019truly}).  
For experiments \kledit{with ground-truth training pairs, the} sample size $K$ of \kledit{the} latent variable $Z$ is set to $5$; for experiments \kledit{with} pairs discovered by UTD, $K$ is set to $10$. 
\kledit{The variance} of $p_\theta(x|z)$ is \hkedit{set to} $0.01$ for both \kledit{pre-training} and correspondence training.
\kledit{For the} KL term annealing, we set \kledit{$k$ to $0.02$ and $s_0$ to $1000$. For CVAE and MCVAE, we set the weight on the KL term to be $0.001$ (weight annealing is used for CVAE). We use the Adam optimizer~\citep{kingma2014adam}} with 
learning rate \hkedit{of} $0.001$. \kledit{The batch size is $100$ for most experiments except those in Section~\ref{sec:unsup}, where we also show results with batch size $400$\jpedit{\footnote{This is mainly for computation considerations. We found that larger batch size lead to a better performance.}}.}

We experiment with data from English and Xitsonga.
English training,
validation and test sets are obtained from the Buckeye corpus~\citep{Pitt2005TheBC}, each with around 6 hours of speech. For Xitsonga, we use \kledit{a 2.5-hour} portion of \hkedit{the} NCHLT~\citep{Barnard2014TheNS} \hkedit{corpus}. \hkedit{The} English \kledit{ground-truth} pairs used in Section~\ref{sec:semi} are generated based on the word \kledit{alignments provided with the Buckeye} corpus. 
Following the same setup as in~\citep{kamper2019truly}, for \hkedit{the unsupervised models in Sections}~\ref{sec:unsup} and~\ref{sec:semi},
we use \hkedit{terms discovered by the UTD system of~\citep{jansen2011efficient}, respectively giving} 10k and 11k segments \hkedit{from the} English and Xitsonga \hkedit{corpora.}
\hkedit{Pairs are generated based on the clusters from the UTD system, resulting in} 14k and 6k pairs\hkedit{, respectively.}
The training, validation and test 
\hkedit{splits} are \hkedit{exactly} the same as in~\citep{kamper2019truly}.  \kledit{The input features are 13-dimensional mel-frequency cepstral coefficients (MFCCs) plus velocity and acceleration vectors.\footnote{While~\cite{kamper2019truly} used only 13-dimensional MFCCs, we found on English development data that all models perform similarly or better when including velocity and acceleration coefficients and we therefore use 39 dimensional MFCCs throughout.}}

\vspace{-.1in}
\subsection{Study of model variants}\label{sec:klAnneal}
\vspace{-.05in}
In this \hkedit{section} 
we \kledit{report on} development experiments to compare options for objective \kledit{functions, as well as disentangle the} effects of the \kledit{pre-training} and correspondence training. 

\begin{figure}
  \centering
  \begin{minipage}{0.475\textwidth}
      \centering
      \includegraphics[width=1\textwidth]{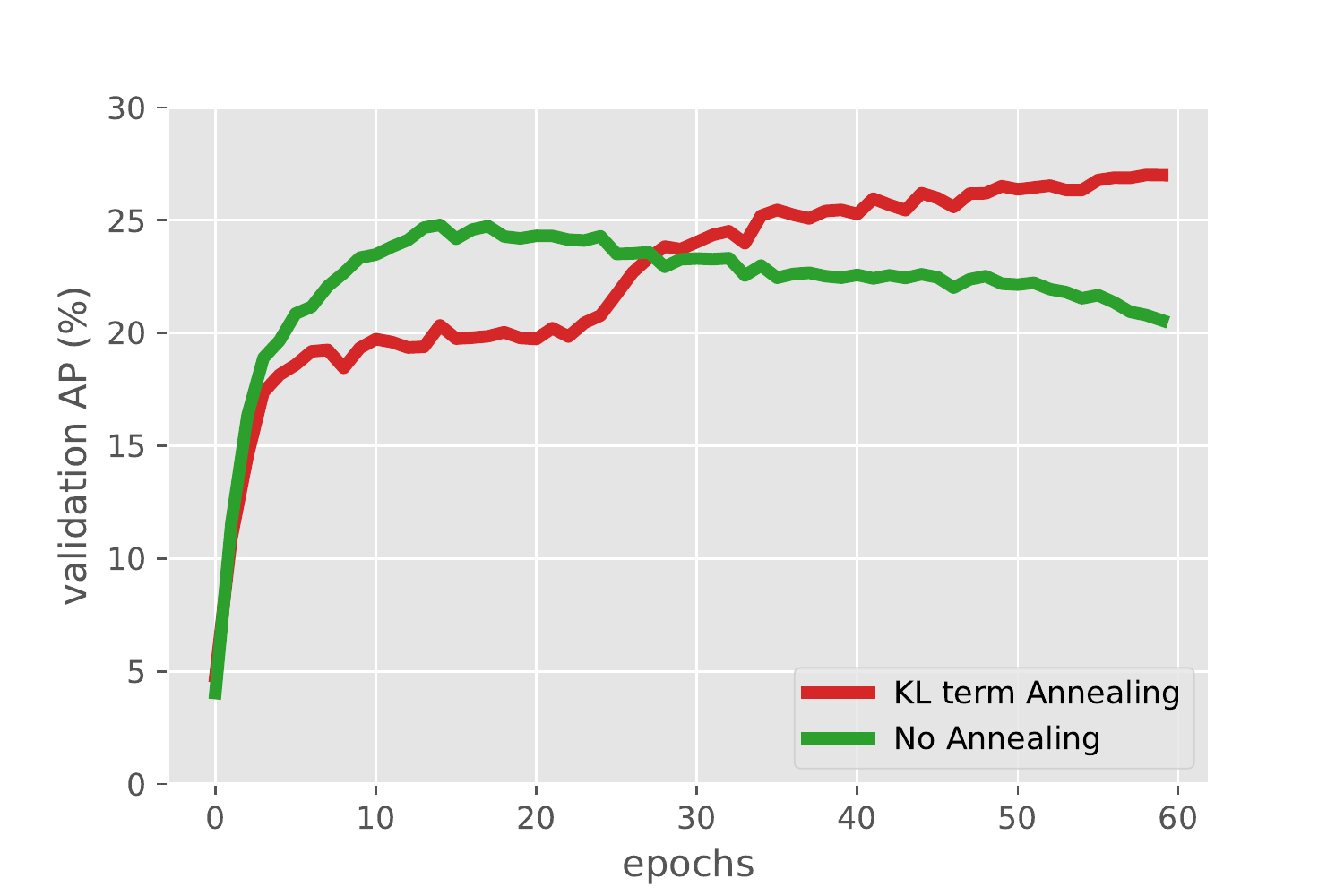} 
      \caption{Comparison of APs between vanilla VAE and VAE with KL term annealed. Both models are trained on ground-truth segments.}\label{fig:preComAps}
  \end{minipage}\hfill
  \begin{minipage}{0.475\textwidth}
      \centering
      \includegraphics[width=1\textwidth]{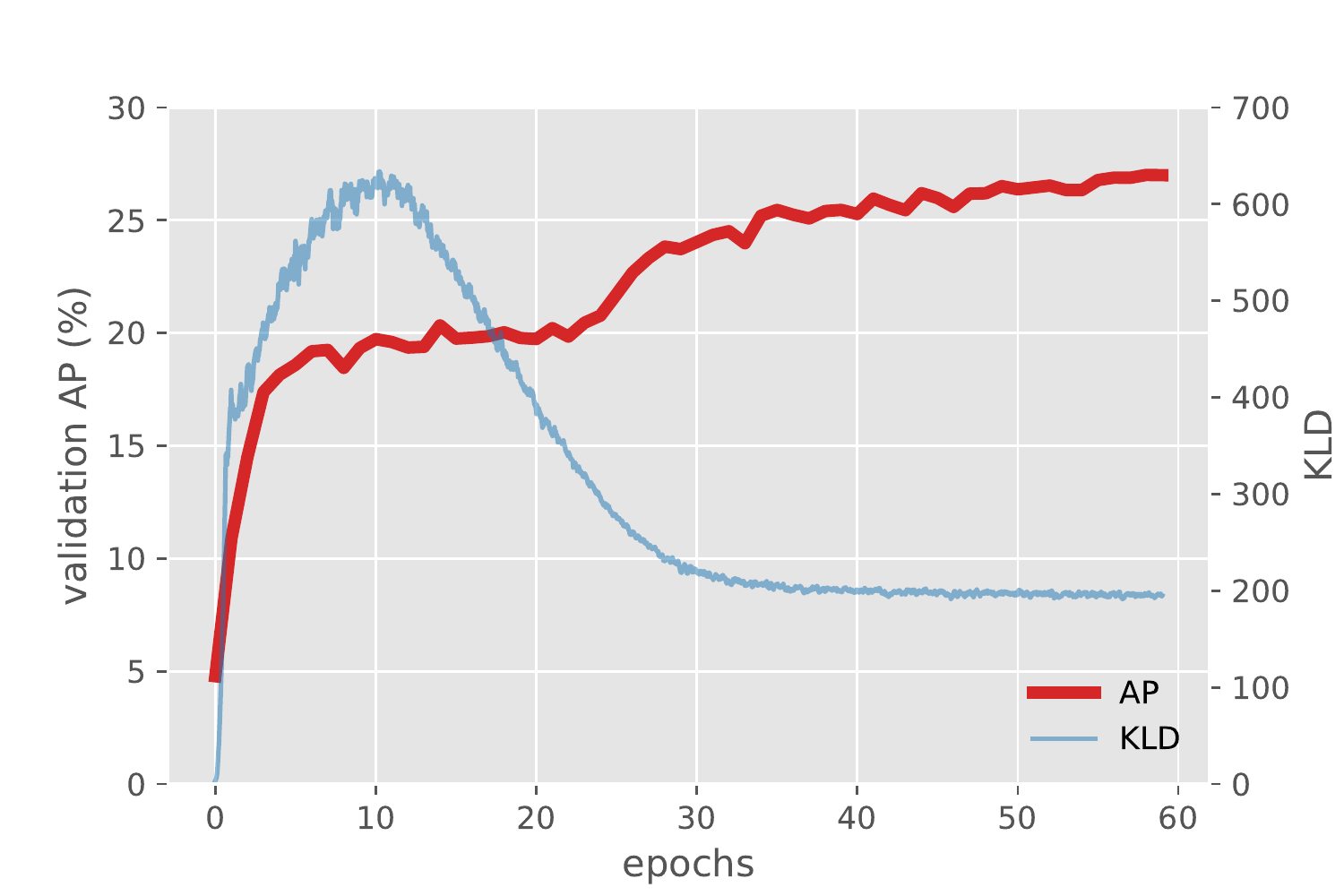} 
      \caption{KL divergence term in the \kledit{VAE} objective plotted alongside the validation AP of the \kledit{VAE with KL term annealing}.}
      \label{fig:preAnneal}
  \end{minipage}
\end{figure}

\jpedit{\textbf{Benefits of \kledit{KL term annealing in pre-training.}}   Figure~\ref{fig:preComAps} is a direct comparison between validation APs (in \kledit{percent}) given by \kledit{a} vanilla VAE and \kledit{a} VAE with the KL term \kledit{annealed,} trained on \kledit{ground-truth} segments. Figure~\ref{fig:preAnneal} shows how the KL term annealing trick helps control the KL divergence while improving the embedding quality.}

\hkedit{\textbf{CVAE vs.\ MCVAE.}}
We train CVAE and MCVAE \kledit{models starting from} the pre-trained \kledit{annealed} VAE. For UTD experiments, all 14k pairs are \kledit{used;} for ground-truth (GT) experiments, 10k pairs are generated based on labels. 
Each experiment is repeated five times with different seeds, and averaged results and standard \kledit{deviations are} reported in Table~\ref{tab:ablation0}.
We see that for both ground-truth pairs and UTD pairs, \hkedit{the} MCVAE achieves 
a better validation AP.

\textbf{Effect of pre-training and correspondence training}. Table~\ref{tab:ablation2} \kledit{shows the separate} and combined \kledit{effects} of pre-training and correspondence training. For both UTD and ground-truth data, combining pre-training and correspondence training gives significantly better results than doing only one of the steps. However, the individual increments \kledit{due to} correspondence training and pre-training are different. For UTD, \kledit{the improvement due to} pre-training is slightly larger than that \kledit{due to correspondence training, whereas} for ground-truth data, \kledit{the improvement due to training is much larger than that due to} pre-training. This indicates the importance of  pair quality in correspondence training. In addition, \kledit{the} pre-trained model on UTD data is slightly better than \kledit{the one pre-trained} on ground-truth data, indicating that pre-training is robust to the quality of the acoustic segments.

\begin{table}[!t]
\vspace{-.2in}
\begin{minipage}[t]{0.475\textwidth}
  \caption{Comparison of \kledit{correspondence VAE-based 
  models} pre-trained as annealed VAEs, \kledit{evaluated on the word discrimination task} on validation data. }\label{tab:ablation0}
  \begin{center}
  \begin{tabular}{llc}
      \toprule
       Data &Model &AP (\%)\\
      \midrule
      UTD&CVAE& 35.4   $\pm$ 0.47\\
      UTD&MCVAE & 37.6  $\pm$ 0.78\\
      \midrule
      GT&CVAE& 55.1 $\pm$ 0.69\\
      GT&MCVAE& 58.8 $\pm$ 0.81\\
      \bottomrule
  \end{tabular}
  \end{center}
  \end{minipage}\hfill
  \begin{minipage}[t]{0.475\textwidth}
  \caption{Ablation study on the effect of correspondence training (`\kledit{corr-train}') and pre-training for \hkedit{the} \kledit{MCVAE, evaluated on the} word discrimination task on validation data.}\label{tab:ablation2}
  \begin{center}
  \begin{tabular}{lccc}
      \toprule
       Data &Pre-train&\kledit{Corr-train} &AP (\%)\\
      \midrule
      UTD&\checkmark& &29.7  $\pm$ 0.49\\
      UTD& &\checkmark&25.6  $\pm$ 0.98\\
      UTD&\checkmark&\checkmark&37.6 $\pm$ 0.78\\
      \midrule
      GT&\checkmark& &27.8  $\pm$ 0.42\\
      GT& &\checkmark&43.6 $\pm$ 1.13\\
      GT&\checkmark&\checkmark&58.8 $\pm$ 0.81\\
      \bottomrule
  \end{tabular}
  \end{center}
  \end{minipage}
  \vspace{-.1in}
  \end{table}

\vspace{-.05in}
\subsection{Test set results for unsupervised acoustic word embeddings}\label{sec:unsup}
\vspace{-.05in}
In this section, we \kledit{present results for models trained} on segments and pairs discovered by UTD \kledit{and evaluated on test data}. 
For \kledit{English, we do early stopping using the 2.7k ground-truth pairs that are available for validation}. 
For Xitsonga, no validation set is available.
\kledit{Instead we use} English UTD data, with size similar to that of \kledit{the} Xitsonga UTD data, to \kledit{choose the number of epochs} for pre-training and correspondence training \kledit{($40$ and $5$ respectively)}. 

Results are reported in Table~\ref{tab:utdcom} \hkedit{for our models together with several previous models and baselines}.
\hkedit{As a naive baseline, downsampling uses 10 equally spaced MFCC vectors from a segment (with appropriate interpolation), and then flattens them to obtain an embedding.
In DTW alignment, the alignment cost of the full sequences is used to make the same-different decision.
We also report results for the}
AE~\citep{kamper2019truly,chung2016unsupervised}, VAE~\citep{kamper2019truly}, \hkedit{annealed} VAE,  SiameseRNN \citep{settle2016discriminative} and CAE-RNN~\citep{kamper2019truly} \hkedit{acoustic word embedding models}.\footnote{\jpedit{We fully tune the network architecture for SiameseRNN, leading to a 2-layer bidirectional GRU network with 400 hidden units. We also fully tune the architecture of CAE-RNN and find that the architecture described in \cite{kamper2019truly} still gives the best validation results. Finally, we also tune the batch size for the two models; unlike for MCVAE, this does not have a large effect on these models, and a batch size of 300 is best for both.}}

Among all models, \hkedit{the} \kledit{MCVAE} performs the best. It not only outperforms its closest AWE \kledit{competitor} (\hkedit{the} CAE-RNN) by 11.3\% and 37.9\% \hkedit{relative} on English and Xitsonga, \kledit{respectively}, but also outperforms DTW, which uses \kledit{the} full uncompressed sequences. To our knowledge, this is the first time an unsupervised acoustic word  embedding  approach  outperforms \kledit{DTW} on the Buckeye corpus.
\hkedit{The} SiameseRNN \kledit{is not competitive} in this \kledit{setting, potentially because it is more reliant on high-quality training pairs than the other models (which is supported by prior work~\citep{Kamper2020ImprovedAW}).}
\begin{table}[!t]
\vspace{-.1in}
  \caption{Word discrimination performance on test data \kledit{for models trained on UTD segments}. 
  }
  \begin{center}
  \begin{tabular}{lcc}
      \toprule
      &\multicolumn{2}{c}{Average Precision (\%)}\\\cmidrule(lr){2-3}
       Model &English&Xitsonga\\
      \midrule
      SiameseRNN&17.5 $\pm$  0.39&25.1 $\pm$ 1.02\\
      AE & 26.4 $\pm$ 0.51&18.0  $\pm$ 0.35 \\
      VAE & 27.7 $\pm$ 0.42&15.7 $\pm$ 0.73 \\
      Annealed VAE (ours) & 29.2 $\pm$ 0.67&17.0  $\pm$ 0.31 \\
      CAE-RNN &35.5 $\pm$ 0.22&32.2 $\pm$ 0.88\\
      MCVAE (ours) & 37.6 $\pm$ 0.65 & 40.2 $\pm$ 0.52\\
      MCVAE (ours, large batch size) & \textbf{39.5 $\pm$ 0.23} & \textbf{44.4 $\pm$ 0.59}\\
      \vspace{-2mm}&\\
      Downsampling~\citep{kamper2019truly} & 21.7\hphantom{\ $\pm$ 0.00} &13.6\hphantom{\ $\pm$ 0.00}\\
      DTW alignment~\citep{kamper2019truly} & 35.9\hphantom{\ $\pm$ 0.00}& 28.1\hphantom{\ $\pm$ 0.00}\\
      \bottomrule
  \end{tabular}
  \end{center}
  \label{tab:utdcom}
  \vspace{-.2in}
  \end{table}
  
\vspace{-.05in}
\subsection{Semi-supervised acoustic word embeddings}\label{sec:semi}
\vspace{-.05in}
In this section we study three AWE \kledit{models---\hkedit{the} CAE-RNN, SiameseRNN and MCVAE---in} a semi-supervised setting, in order to mimic the low-resource language processing scenario.
In this setting, we \kledit{start with models from the previous section that have been trained in an unsupervised way, and continue training them but now using}
ground-truth pairs. To show how different amounts of ground-truth data improve the results, we increase the \kledit{number of} training pairs from only 1k to 50k.

\begin{figure}[h]
\vspace{-.2in}
  \begin{center}
    \includegraphics[width=0.6\textwidth]{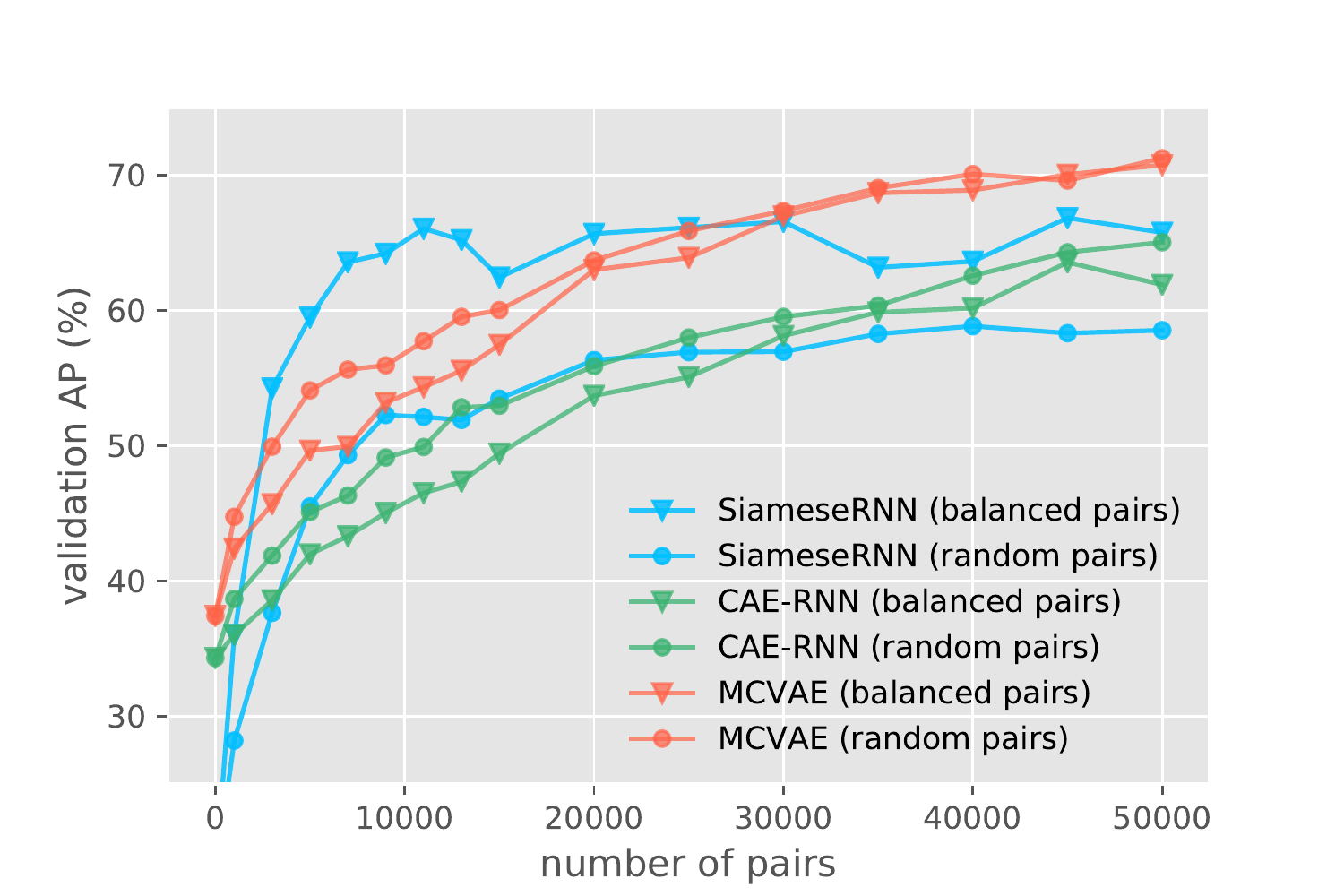}
    \caption{Validation AP vs.~number of ground-truth pairs \kledit{for supervised training, 
    for models pre-trained on UTD data.}}\label{fig:semi}
  \end{center}
  \vspace{-.25in}
\end{figure}

To \kledit{examine the} models' performance under different training data distributions, \kledit{we generate pairs in two} ways: (1) \kledit{\bf Random pairs:} randomly sample from the dataset, and thus the distribution of pairs will follow the word frequency \kledit{distribution} in the corpus.
This will lead to an imbalanced training set, but the data distribution will be \kledit{similar to that of the} validation set. 
(2) \kledit{\bf Balanced pairs:} generate pairs \kledit{with a more balanced distribution.} 
To achieve this, we set an upper bound for \kledit{the} number of segments of \kledit{the} same word type based on the total number of pairs we want to generate. For example, to get 25k pairs in total, we 
generate at most 5 pairs \kledit{of each word}. 
The result is shown in Figure~\ref{fig:semi}.

There are several interesting things to notice in Figure~\ref{fig:semi}. 
First, 
\hkedit{the}
\kledit{MCVAE} is the most \kledit{data-efficient} model within the range of 
\kledit{data sizes} that we consider here, as \kledit{most} of the time \hkedit{the} MCVAE outperforms other models by a large margin. 
\kledit{Second, the encoder-decoder} based models (\hkedit{the} MCVAE and CAE-RNN) \kledit{are} more robust \kledit{to changes in the training data distribution than the discriminative} SiameseRNN. Among the two encoder-decoder models, \hkedit{the} MCVAE is more robust than \hkedit{the} CAE-RNN, as 
\jpedit{gap between between the random and balanced training set results is narrowed as the amount of training data increases.}
Third, \kledit{the} SiameseRNN \hkedit{performs} 
surprisingly well in the low to median data regime (3k - 25k) when training pairs are balanced; however, it reaches its best performance at 11k and plateaus afterwards. 

\vspace{-.05in}
\section{\kledit{Conclusion}}
\vspace{-.05in}
We \kledit{have} presented the MCVAE, a generative model for unsupervised and semi-supervised \kledit{learning of acoustic word embeddings\kledit{. We have shown} that \hkedit{it} is robust to \hkedit{the} amount, distribution, and quality of training data.} On \hkedit{the English} Buckeye corpus, this is the first time that an unsupervised acoustic word embedding approach outperforms \kledit{DTW}. 
Future work includes explicitly incorporating label information for training when available \kledit{(as in, e.g.,~\cite{He2017MultiviewRN,Bengio2014WordEF})}, generating higher quality segment pairs for unsupervised training, and applying the model in downstream tasks such as 
\hkedit{query-by-example} 
speech search.

\bibliographystyle{abbrvnat}
\bibliography{mcvae}
\end{document}